%% file: main.tex
\documentclass[sigconf]{acmart}
\usepackage{caption}
\usepackage{url}
\usepackage{subcaption}
\usepackage{graphicx}
\usepackage{algorithm}
\usepackage{algorithmic}
\usepackage{xspace}

\newcommand{\ie}{{\em i.e.},\xspace}
\newcommand{\eg}{{\em e.g.},\xspace}

\newcommand{\modify}[1]{\textcolor{black}{#1}}

\newcommand{\methodname}{Hue\xspace}

\acmConference[ESEC/FSE 2023]{The 31st ACM Joint European Software Engineering Conference and Symposium on the Foundations of Software Engineering}{11 - 17 November, 2023}{San Francisco, USA}

\AtBeginDocument{%
  \providecommand\BibTeX{{%
    \normalfont B\kern-0.5em{\scshape i\kern-0.25em b}\kern-0.8em\TeX}}}

\setcopyright{acmcopyright}
\copyrightyear{2023}
\acmYear{2023}
\acmDOI{XXXXXXX.XXXXXXX}

\acmPrice{15.00}
\acmISBN{978-X-XXXX-XXXX-x/YY/MM}

\begin{document}

\title{Hue: A User-Adaptive Parser for Hybrid Logs}

\author{Junjielong Xu}
\email{siyuexi@foxmail.com}
\affiliation{%
    \institution{The Chinese University of Hong Kong, Shenzhen}
    \country{China}
}
\author{Qiuai Fu}
\email{fuqiuai@huawei.com}
\affiliation{%
    \institution{Huawei Cloud Computing Technologies CO., LTD.}
    \country{China}
}
\author{Zhouruixing Zhu}
\email{zhuzhouruixing@icloud.com}
\affiliation{%
    \institution{The Chinese University of Hong Kong, Shenzhen}
    \country{China}
}
\author{Yutong Cheng}
\email{222010519@link.cuhk.edu.cn}
\affiliation{%
    \institution{The Chinese University of Hong Kong, Shenzhen}
    \country{China}
}
\author{Zhijing Li}
\email{zhijingbaby@gmail.com}
\affiliation{%
    \institution{The Chinese University of Hong Kong, Shenzhen}
    \country{China}
}
\author{Yuchi Ma}
\email{mayuchi1@huawei.com}
\affiliation{%
    \institution{Huawei Cloud Computing Technologies CO., LTD.}
    \country{China}
}
\author{Pinjia He*\let\thefootnote\relax\footnotetext{*Corresponding author}}
\email{hepinjia@cuhk.edu.cn}
\affiliation{%
    \institution{The Chinese University of Hong Kong, Shenzhen}
    \country{China}
}

\setcopyright{none}
\settopmatter{printacmref=true} 
\renewcommand\footnotetextcopyrightpermission[1]{} 

\renewcommand{\shortauthors}{Xu et al.}

\begin{abstract}
Log parsing, which extracts log templates from semi-structured logs and produces structured logs, is the first and the most critical step in automated log analysis.
While existing log parsers have achieved decent results, they suffer from two major limitations by design.
First, they do not natively support hybrid logs that consist of both single-line logs and multi-line logs (\eg Java Exception and Hadoop Counters). 
Second, they fall short in integrating domain knowledge in parsing, making it hard to identify ambiguous tokens in logs.
This paper defines a new research problem, \textit{hybrid log parsing}, as a superset of traditional log parsing tasks, and proposes \textit{\methodname}, the first attempt for hybrid log parsing via a user-adaptive manner.
Specifically, \methodname converts each log message to a sequence of special wildcards using a key casting table and determines the log types via line aggregating and pattern extracting.
In addition, Hue can effectively utilize user feedback via a novel merge-reject strategy, making it possible to quickly adapt to complex and changing log templates.
We evaluated Hue on three hybrid log datasets and sixteen widely-used single-line log datasets (\ie Loghub).
The results show that \methodname achieves an average grouping accuracy of 0.845 on hybrid logs, which largely outperforms the best results (0.563 on average) obtained by existing parsers.
\methodname also exhibits SOTA performance on single-line log datasets.
Furthermore, \methodname has been successfully deployed in a real production environment for daily hybrid log parsing.

\end{abstract}





\maketitle

\input{Sections/1_Introduction}
\input{Sections/2_Motivation}

\input{Sections/3_Definition}

\input{Sections/4_Approach}

\input{Sections/5_Experiments}

\input{Sections/6_Industrial}

\input{Sections/7_Threats}
\input{Sections/8_Related}
\input{Sections/9_Conclusion}

\section{Data Availability}
We uploaded our repository at \url{https://github.com/logpai/hybridlog}

\balance
\bibliographystyle{ACM-Reference-Format}
\bibliography{ref}

\end{document}

%% file: Sections/1_Introduction.tex
\section{Introduction}\label{sec:intro}

In recent years, software systems, such as online services (\eg Google Search and Bing Search) and system software (\eg Android and Windows), have become an integral part of our daily lives, which generate extremely large amounts of software logs every day.
These logs can be used in various tasks, \eg anomaly detection~\cite{deeplog, exp, mininglog, robustlog}, root cause analysis~\cite{localizelog, idlogm, idlogc}, failure prediction~\cite{prefix}, log compression~\cite{logzip, feas}, and user profile construction~\cite{derive}.
Because of the rapid growth of the log volume, it is difficult to identify valuable information from the massive log data manually. 
To this end, automatic log analysis has been widely studied in recent years. 
The first step of automatic log analysis is log parsing, which aims at extracting log templates and converting semi-structured log messages into structured log messages for downstream tasks. 
Specifically, the core task of log parsing is to distinguish between \textit{constants} and \textit{variables} in log messages, where constants are the tokens written by developers in the logging statements (\eg a description of a software operation) and variables are tokens that can change according to runtime environments (\eg an IP address).
Recent log parsers~\cite{ael,lenma,spell,iplom,drain,spine} have achieved decent results on open-source log datasets (\ie Loghub~\cite{benchmark}).
However, they still suffer from two major limitations:

\begin{figure}[tbp]
\centering
\includegraphics[scale=0.7]{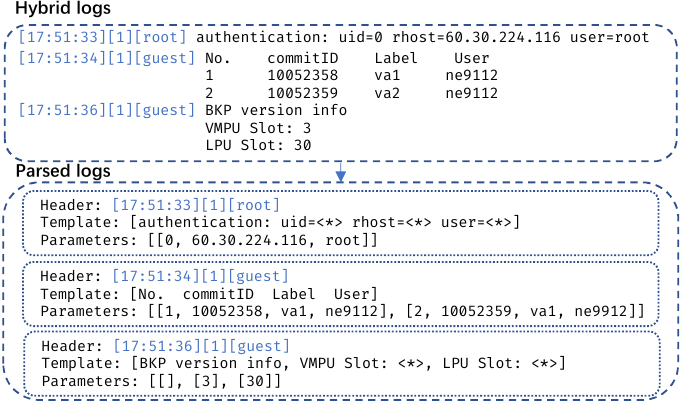}   
\caption{A simplified example of hybrid log parsing.}
\label{fig:parsingexample}
\end{figure}

\textit{First}, existing parsers assume the incoming logs are single-line, and they try to extract the common pattern as templates line by line.
However, in practice, log messages could be multi-line, such as KPI tabular echos (\eg CPU usage), tracebacks (\eg Java Exception), and key-value pairs (\eg Hadoop Counters).
In addition, due to the complexity of modern software and the centralized log collection practice, software logs could be \textit{hybrid}, where single-line log messages mix with multi-line log messages.
For example, Fig.~\ref{fig:parsingexample} presents an example of hybrid log parsing that contains a single-line log message and two multi-line log messages.
In our opinion, hybrid log parsing is a more general research problem that aims to parse either single-line logs, multi-line logs, or their mix, while recent research mainly focuses on parsing single-line logs. 

\textit{Second}, since the logging statements of these log messages are often inaccessible (\eg written in third-party libraries), it is difficult to tell whether a token is a constant or a variable. 
People could defines different templates for the same logs due to the disagreement on some ambiguous tokens. 
For example, recent research~\cite{uniparser, guideline} reports that the widely-used parsing datasets Loghub contains labeling errors, which are typically caused by the difference in opinions.
In addition, whether a log template is "correct" or not depends on the requirements of the downstream tasks.
For example, in a real business scenario in Huawei Cloud, the developer-labeled log template for the console log \texttt{"display ipv6"} is \texttt{"<*> ipv6"} in the root cause analysis task and \texttt{"display <*>"} in the user profile construction task.
In practice, many "correct" log templates can be summarized only with expert domain knowledge.
Thus, we argue that human feedback is "the last mile" in log parsing and a log parser that can effectively and efficiently integrate human feedback is highly in demand.

In this paper, we introduce a new, general, and practical research problem, \textit{hybrid log parsing}, which aims to parse single-line logs, multi-line logs, and their mix.
To this end, we propose \textit{\methodname}, the first hybrid log parser that works in an online manner and efficiently adopts human feedback.
\methodname is simple yet effective.
\methodname adopts key casting, line aggregating, and pattern extracting to precisely parse complex hybrid logs.
We further design a novel mechanism that allows users to reject a template update on potentially ambiguous tokens.
We evaluate our approach on (1) three hybrid log datasets collected from both open-source software and Huawei Cloud's cloud services, and (2) sixteen widely-used single-line log datasets (\ie Loghub).
The results show that \methodname achieves an average grouping accuracy (GA) of 0.845, which largely outperforms the best result obtained by existing parsers (0.563 on average). 
\methodname also exhibits SOTA accuracy on single-line log parsing without leveraging the human feedback component, achieving the highest GA on 8 out of 16 datasets, which is the most among all compared parsers.
\methodname has been successfully deployed in the cloud services in Huawei Cloud to parse daily hybrid logs.



This paper makes the following main contributions:
\begin{itemize}
    \item It introduces a new, general, and practical research problem, hybrid log parsing, which is a superset of the existing single-line log parsing problem.
        \item It proposes \methodname, the first log parser that natively supports hybrid log parsing and has a human feedback integration mechanism that largely reduces unnecessary queries.
        \item It presents the evaluation of \methodname using three hybrid log datasets and sixteen single-line log datasets, demonstrating that \methodname achieves SOTA accuracy and efficiency on both hybrid and single-line logs.
        \item It releases two hybrid log datasets collected from open-source software and cloud system for this research direction.
\end{itemize}





%% file: Sections/2_Motivation.tex
\section{Motivation}\label{motivation}


This section intends to further explain the two major limitations of existing parsers that motivates our work. 
The following is mainly inspired by and summarized from our collaboration experience with engineers in Huawei Cloud.


\subsection{Neglected Hybrid Logs}\label{sec:neglected}

\begin{figure}[tbp]
 
\centering
\includegraphics[scale=0.6]{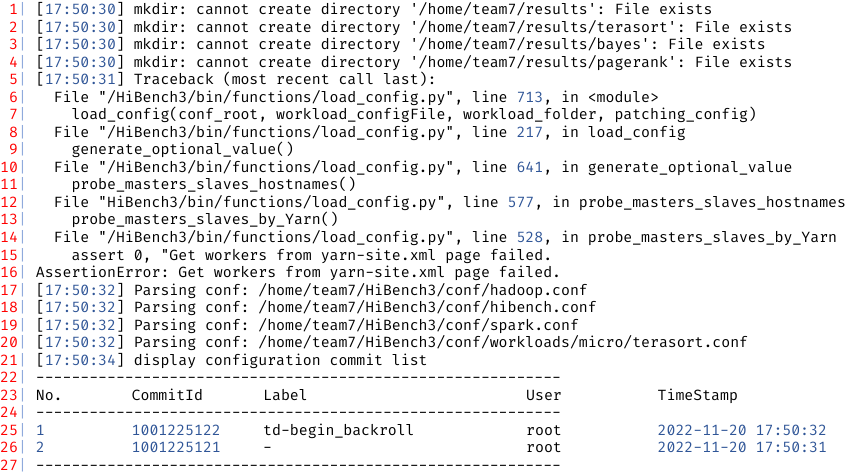}   
\caption{An example of hybrid logs collected from an industrial log management platform.}
\label{fig:hybridlogexample}
\label{sec2_1}
 
\end{figure}

Hybrid logs are a common type of data in the software industry. 
As shown in Fig.~\ref{fig:hybridlogexample}, they are usually generated from cloud platforms and can contain a mixed combination of single-line and multi-line log messages. 
The hybrid nature of the logs is often caused by permission restrictions between different service departments within an organization.
In particular, IT operators may not be able to access logs in all components directly and must instead use a log aggregation system to gather all the service outputs for centralized management. 
This means that hybrid logs may contain a variety of log types, including single-line logs such as component event logs and multi-line logs such as console echoes, exception traceback, and even tabular system key performance indicators (KPIs).
Hybrid logs are not only found in closed-source software, but they are also commonly encountered in open-source components such as Hadoop, Spark, and MySQL. 
This highlights the widespread availability of hybrid logs in both closed-source and open-source software.



Hybrid logs contain valuable information that can be utilized in many downstream tasks, including IT operators using the traceback for failure prediction and root cause analysis, QA engineers selectively testing APIs through KPI tables, and commercialization departments constructing user profiles from console echoes to target potential user needs. 
However, the structural variability of hybrid logs presents a challenge for current log parsers, leading to them ignoring these logs during training and deployment. 
If single-line parsers are used in pipelines, they tend to split a multi-line log message into multiple lines of text and regard each line as a separate single-line log message, leading to incorrect parsing results.
Even if IT operators try to use special regex to flatten hybrid logs, tabular messages may still not be parsed correctly because of the uncertainty in the number of table rows.
To address these issues, this paper proposes key casting (Sec.~\ref{sec:keycast}) and line aggregating techniques (Sec.~\ref{sec:lineaggregate}). 
The limitations of existing work are further discussed in our experiments (Sec.~\ref{sec:rq1}).


\subsection{Ambiguous Template Tokens}\label{sec:ambiguous}

\begin{figure}[tbp]
 
\centering
\includegraphics[scale=0.62]{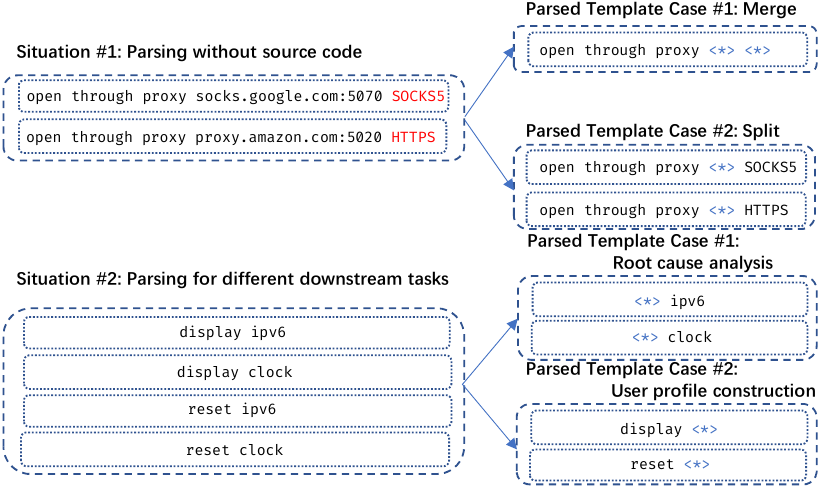}   
\caption{Two typical situations that difficult for the parser to determine the template automatically.}
\label{fig:ambiguousexample}
\end{figure}

Recent research found that different people could generate different log templates because of ambiguous tokens~\cite{uniparser, guideline}. 
In Huawei Cloud, there are two typical situations that confuse parsers. 
(1) Whether to merge two candidate templates by turning a constant into a variable. 
For example, in Fig.~\ref{fig:ambiguousexample}, Situation 1, it is debatable whether the two logs belong to the same event (and should be merged). 
(2) A log message could have multiple "correct" templates when it was used by different downstream tasks. 
For example, in Fig.~\ref{fig:ambiguousexample}, Situation 2, for user profiling modeling, the analyst cares about the instructions and actions the user performed, whereas for root cause analysis, the analyst cares about the object being manipulated. 
The difference in their preferences leads to different "correct" log templates.




To improve the compatibility and generalizability of the log templates generated in log parsing, we introduce an efficient manual feedback mechanism (Sec.~\ref{sec:updating}) that works seamlessly within a parse tree and effectively improves parsing accuracy with little manual involvement. 
The benefits of this mechanism are further explained in our experiments (Sec.~\ref{sec:rq3}).


%% file: Sections/3_Definition.tex
\section{Preliminary}

This section intends to introduce the main concepts in hybrid log parsing, especially our definition and the characteristics of \textit{hybrid log} and the input and output of \textit{hybrid log parsing}.


\subsection{Hybrid Logs}\label{sec:hybridlogs}

Hybrid logs could be single-line logs, multi-line logs, or their mix.
Based on our experience with hybrid log analysis in Huawei Cloud, in this paper, we mainly consider three kinds of logs in hybrid logs: \textit{event logs} (single-line), \textit{table logs} (multi-line), and \textit{text logs} (multi-line).
For example, in Fig.~\ref{fig:hybridlogexample}, line 1-4 and line 17-20 are event logs, line 5-16 is a text log, and line 21-27 is a table log.


\subsubsection{Event Log}

Event logs are single-line log messages.
An event log records an operation or status of a service or component and contains a log header and message content.
To align with the definition in existing single-line log parsing research~\cite{guideline, survey}, an event log's template consists of constants in the log message and wildcards that indicate variables.


\subsubsection{Table Log}
Table logs are multi-line logs that contain a table header, multiple lines of parameters, and potentially table lines.
Table logs with the same log template (1) have the same number of table columns and the token data type in the same column is the same (\eg the timestamps in the last column in the table log in Fig.~\ref{fig:hybridlogexample}); (2) might have different numbers of rows.
Typical table logs contain various system metrics and these logs are often generated by the performance testing component in the cloud and the echoes from the user shell. 
Since the header of the table logs consists of the key information (\ie column number and column types) that distinguishes between different tables, the template of a table log is the table header, while the table content is regarded as parameters in the parsed log message (as illustrated in Fig.~\ref{fig:parsingexample}).
Therefore the goal of log parsing on table logs is to extract its table header and transform its table content into parameters.


\subsubsection{Text Log}

Text logs are multi-line logs that do not fall into the "table log" category.
Text logs record the detailed information in addition to the system event in plain text, such as traceback call stack in a program or database content in the form of key-value pairs. 
Text logs are widely available in many components such as Hadoop, Spark, and MySQL.
However, to the best of our knowledge, only a little log analysis research utilized this kind of log and we think one of the main reasons is that existing log parsers cannot accurately parse text logs.
We define the template of a text log as tokens in several lines that remain constant, such as the traceback error type or the keys in the message. 
Thus, the goal of log parsing on text logs is to extract these lines from the message.


\subsection{Hybrid Log Parsing}\label{sec:hybridparsing}



The goal of hybrid log parsing is to extract log templates from log messages and generated structured logs accordingly. 
In our opinion, hybrid log parsing, which focuses on single-line logs, multi-line logs, and their mix, is a \textit{superset} of existing log parsing research that targets single-line logs.
The output of this task includes (1) a meta-file that contains the template of each log message and its location in the original log file, and (2) files with structured variable lists (\eg headers and parameters) in each log type.

In this paper, we view hybrid log parsing as a classification problem.
Given a raw log message $L$ consisting of $n$ tokens across $l$ lines, denoted as $[t_1, t_2, ..., t_n]$,
a hybrid log parser should predict the type $T$ of log message $L$,
the log template group $g_k^T \in G^T=\{g^T_1, g^T_2, ...,g^T_N\}$ it belongs to among the total $N$ templates in type $T$,
and whether a token $t_i$ is a constant $(y_i=1)$ or a variable $(y_i=0)$.
Finally, the log parser must output its predictions in a structured format.
Hybrid log parsing is challenging because it is non-trivial to automatically distinguish between different types of logs when they are mixed and identify the constants and variables in them.


%% file: Sections/4_Approach.tex
\section{Approach}
\begin{figure*}[htbp]
 
\centering
\includegraphics[scale=1.42]{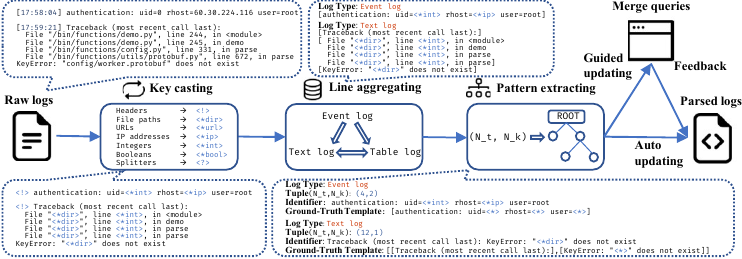}   
\caption{
The overview of \methodname \modify{with two examples}. 
Users can customize the key casting table and provide feedback to optimize \methodname for more complex data types and usage scenarios based on the target data characteristics and their own domain expertise.
}
\label{fig:flowexample}
\end{figure*}

\subsection{Overview}\label{sec:overview}

This section introduces \methodname, an unsupervised online log parser that utilizes a heuristic parse tree and an efficient human feedback integration mechanism. 
The overview of \methodname is presented in Fig.~\ref{fig:flowexample}, 
which contains four main components: key casting, line aggregating, pattern extracting, and \modify{online updating}. 

\modify{Specifically, \methodname streamly reads the log data line by line, splits each line into tokens by spaces, and sends the raw \textit{token sequence} to main components for parsing:} 
\modify{First, \textit{key casting} transforms part of the tokens into \textit{keys}, \ie special wildcards indicating representative token data type.
For instance, a raw token sequence with $n$ tokens $S_{raw}=[t_1, t_2, t_3, ..., t_n]$ can be casted to a new token sequence $S = [t_1, \mathcal{K}_2, \mathcal{K}_3, ..., t_n]$, where each $\mathcal{K}$ represents a specific key.}
\modify{When all lines in a log are fully casted, \textit{line aggregating} combines adjacent token sequences into \textit{blocks} (\ie lists of adjacent token sequence) via heuristic rules.
For example, a log with $l$ lines $L = [S_1, S_2, ..., S_l]$ can be aggregated to $L' = [B_1, B_2, ..., B_k] = [[S_1], [S_2, S_3], ..., [S_l]]$, which contains $k$ blocks.}
\modify{Then, \textit{pattern extracting} ultilizes aggregated blocks to obtain log tempates.}
\modify{Finally, \modify{\textit{online parsing}} revises the current log template in an online manner with the guidance of an optional human feedback mechanism.}

Note that hybrid logs could be single-line logs, multi-line logs, or their mix, thus \methodname, which is the first hybrid log parser, can be utilized to parse single-line logs as existing parsers.

\subsection{Key Casting}\label{sec:keycast}

Preprocessing has been widely adopted by existing parsers.
A typical preprocessing process identifies common variables (\eg IP address) by regular expressions, and then either remove the identified tokens or replaces them with the wildcard \texttt{"<*>"}~\cite{survey}. 
However, these preprocessing strategies are not effective when dealing with hybrid logs. 
which may contain table logs and text logs.
Table logs have lines with only variables, which would be all removed or transformed into the same wildcard \texttt{"<*>"} by existing preprocessing strategies, leading to significant inaccuracy.
In addition, these strategies might exacerbate the difficulty of distinguishing text logs from table logs.
To address this problem, it is possible to manually configure a list of delimiters for each system separately.
However, this is impractical for hybrid logs because they are often collected from different system components or even different systems.

\methodname adopts a novel preprocessing strategy called \textit{key casting}.
The main idea is to use different wildcards (\eg \texttt{"<*int>"} and \texttt{"<*ip>"}) to better encodes prior knowledge in preprocessing. 
\methodname split a line by spaces and transforms tokens that have been commonly used as variables (\eg IP addresses, file paths, boolean values) into the corresponding wildcards called "keys" according to a general casting table. 
For example, in Fig.~\ref{fig:flowexample}, \texttt{"authentication: uid=0 rhost=60.30.224.116 user=root"} would be transformed into \texttt{"authentication: uid=<*int> rhost=<*ip> user=root"}. 
Key casting is simple yet effective for hybrid log parsing, and we believe it could also benefit existing log parsing approaches as a general preprocessing strategy. 
In our implementation, we construct a general key casting table for all datasets, which contains seven kinds of keys, as illustrated in Fig.~\ref{fig:flowexample}.
Users can also easily customize their own key casting table by changing one line of code.

\begin{figure}[tpb]
 \centering
\includegraphics[scale=0.7]{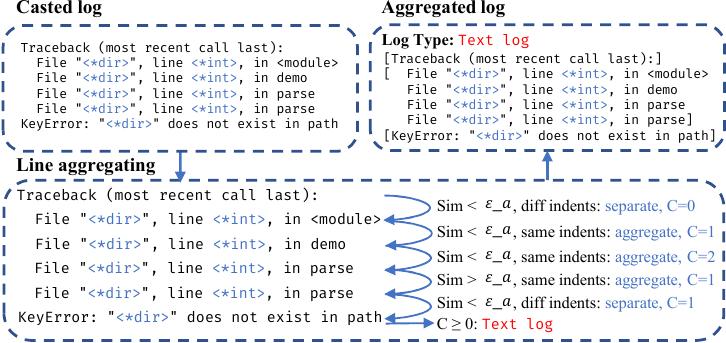}    
\caption{
\modify{
An example of line aggregating. (assume $\varepsilon_a = 0.9$)
}
}
\label{fig:aggregatingdetails}
\end{figure}

\subsection{Line Aggregating}\label{sec:lineaggregate}


\modify{To correctly parse hybrid logs, identifying log type is critical.
Intuitively, using line numbers to identify event logs is relatively easy; but distinguishing between table logs and text logs is challenging because both of them have a multi-line structure.
To overcome this issue, we propose \textit{line aggregating} to differentiate between table logs and text logs.
Specifically, we observe that adjacent lines in table logs tend to exhibit higher similarity, whereas in text logs, although adjacent lines are less similar, they often share the same indentation.
Thus, \methodname aggregates adjacent lines in multi-line logs and identifies log types based on adjacent lines' (1) sequence similarity and (2) indent number.}

\modify{The workflow of line aggregating is shown in Algo.~\ref{alg:1}.
\methodname first determines whether the current log is an event log by checking if it only consists of single line (line 2). 
If the log has multiple lines, \methodname sequentially aggregates adjacent lines into several \textit{blocks} (Sec.~\ref{sec:overview}) if sequence similarity exceeds the aggregating similarity threshold $\varepsilon_a$ (\ie a hyperparameter determined before parsing) or they have the same indentation (line 4-13). 
Then, \methodname utilizes a \textit{type counter} $C$ to keep track of the predominant reason in the aggregation process (line 14): if $C < 0$, sequence similarity is the primary reason for aggregating, suggesting that the log should be a table log; if $C \geq 0$, indentation number is the primary reason for aggregating, suggesting that the log should be a text log. 
Finally, \methodname utilizes the aggregated logs for subsequent pattern extraction.
To enhance understandability, we also provide an example of line aggregating for text logs in Fig.~\ref{fig:aggregatingdetails}.
Specifically, similarity $Sim(S_1,S_2)$ represents the proportion of identical tokens at the same positions in two sequences (Eq.~\ref{eq:1}), where $N$ is their average token length, and $Equ(t_1,t_2)$ measures the equality of two tokens (Eq.~\ref{eq:2}).}
\modify{
\begin{equation}\label{eq:1}
    Sim(S_1,S_2) = \frac{\sum_{i=1}^NEqu(S_1[i],S_2[i])}{N}
\end{equation}
}
\modify{
\begin{equation}\label{eq:2}
    Equ(t_1,t_2) =
    \left
    \{
    \begin{aligned}
    & 1 &\text{if $t_1=t_2$} \\
    & 0 &\text{if $t_1\neq t_2$}
    \end{aligned}
    \right.
\end{equation}
}
\modify{}

\begin{algorithm}[tpb]
	\renewcommand{\algorithmicrequire}{\textbf{Input:}}
	\renewcommand{\algorithmicensure}{\textbf{Output:}}
	\caption{Line Aggregation \& Type Determination}
    \label{alg:1}
	\begin{algorithmic}[1]
    \REQUIRE Sequence queue of current message, $\mathcal{Q}=\{S_1,S_2,....,S_n\}$\\ \ \ \ \ \ \
    \modify{Aggregating similarity threshold $\varepsilon_a$}
    \STATE \modify{Log type $\mathcal{T}=\text{"\texttt{EVENT}"}$, Block queue $\mathcal{Q_B}=\{\}$, Last block $\mathcal{B}_{last} = S_1$}
    \IF{$|\mathcal{Q}|>1$}
    \STATE Type counter $\mathcal{C}=0$
    \FOR{$S_i \in \mathcal{{Q}}$}
    \STATE \modify{$S_{B_{last}} = $ the common sequence of all $S \in \mathcal{B}_{last}$}
    \IF{$Sim(S_i, S_{B_{last}}) \ge $\modify{$\varepsilon_a$}}
    \STATE $\mathcal{C} \leftarrow \mathcal{C}-1$, $\mathcal{B}_{last} \leftarrow \mathcal{B}_{last} \cup S_i$
    \ELSIF{$S_i$ and $S_{B_{last}}$ have the same indents} 
    \STATE $\mathcal{C} \leftarrow \mathcal{C}+1$, $\mathcal{B}_{last} \leftarrow \mathcal{B}_{last} \cup S_i$
    \ELSE 
    \STATE $\mathcal{Q_B} \leftarrow \mathcal{Q_B} \cup \mathcal{B}_{last}$, $\mathcal{B}_{last} = \{S_i\}$
    \ENDIF
    \ENDFOR
    \STATE \modify{$\mathcal{T}=\text{"\texttt{TABLE}"}$ if $\mathcal{C} < 0$, else: $\mathcal{T}=\text{"\texttt{TEXT}"}$}
    \ENDIF
    \STATE $\mathcal{Q_B} \leftarrow \mathcal{Q_B} \cup \mathcal{B}_{last}$
    \ENSURE $\mathcal{Q_B}$, $\mathcal{T}$
	\end{algorithmic} 
\end{algorithm}

\modify{Notably, there are a few details in implementing line aggregating:}
(1) Lines consisting of only delimiters or null characters should be skipped. 
(2) Based on our observation in both open-source and industrial software, hybrid logs are often preceded by log headers, which can serve as delimiters between logs. 

\subsection{Pattern Extracting}\label{sec:patternextract}

To extract log templates, log parsers often first group logs with the same template into the same group~\cite{drain,spine,lenma}. 
Traditional log parsers often use complete log messages (all tokens) for clustering, which is inefficient in hybrid log parsing (shown in Fig.~\ref{fig:exectimehybrid}) because a hybrid log message can be very long after flattened. 
In addition, the line number of multi-line logs from the same template group can vary greatly, leading to low accuracy in message-level clustering.

\begin{figure}[tpb]
 \centering
\includegraphics[scale=0.75]{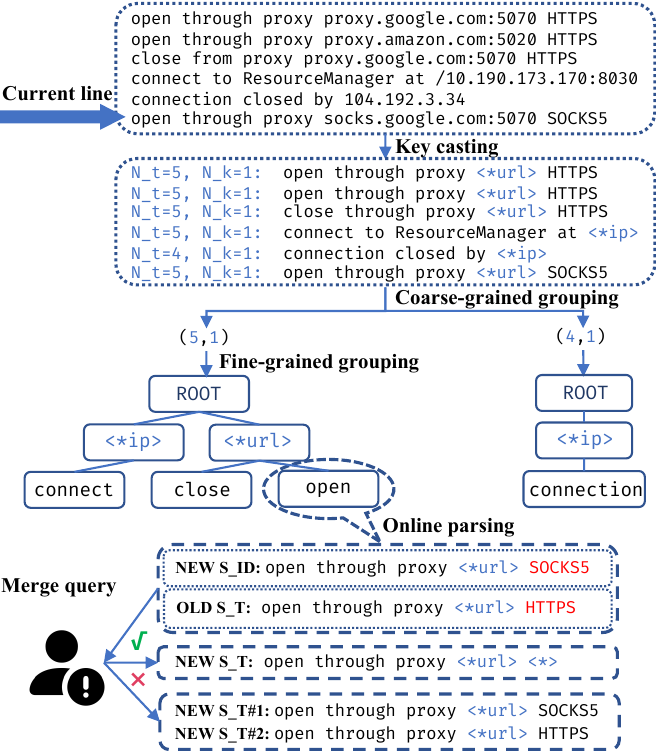}    
\caption{
\modify{
An example of pattern extracting and guided online parsing for event logs. 
(assume maximum tree depth is 2) 
}
}
\label{fig:flowdetail}
\end{figure}

To tackle these challenges, \methodname proposed \textit{pattern extracting}, which utilizes the block queue $Q_B$ and log type $T$ obtained from line aggregating to group logs with the same templates.
As shown in Fig.~\ref{fig:flowexample}, \methodname uses a special sequence of each log, the "log identifier" ($S_{ID}$), which may contain a log template as defined in Sec.~\ref{motivation}, as a manual feature.
\methodname assumes that logs from the same template have similar $S_{ID}$, and uses this information to group logs.
Specifically, $S_{ID}$ is (1) the entire sequence for an event log, (2) the concatenated common sequences from the first and last blocks for a text log, and (3) the sequence of the first block with a capacity of one before the first block with a capacity greater than one for a table log.
The choice of the identifier is based on an assumption that text logs often have template tokens at the beginning or end (\eg Error Type), and that table headers are typically distinct and should be aggregated with the table content.
4 as shown in Fig.\ref{fig:flowdetail}.

\begin{figure*}[htbp]
 \centering
\includegraphics[scale=0.8]{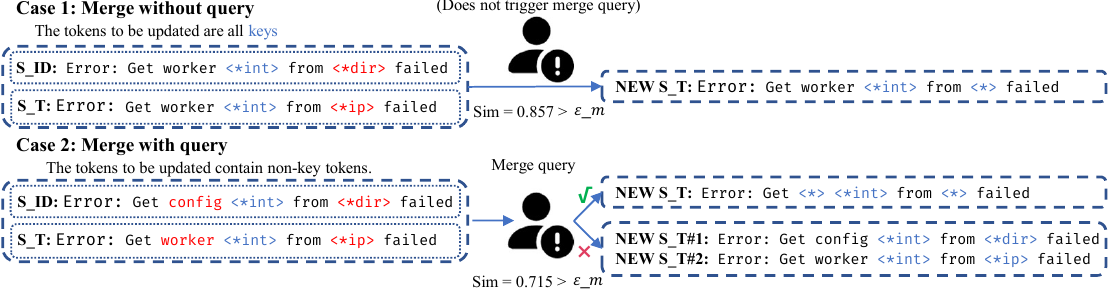}    
\caption{
\modify{
Two cases of guided updating mode. The updates of keys do not need human guidance since they are typically considered as non-ambiguous tokens.
(assume $\varepsilon_m = 0.7$)
}
}
\label{fig:feedbackdetails}
\end{figure*}

\subsubsection{Coarse-grained Grouping}

To group logs, \methodname counts the number of tokens and keys in each log's identifier $S_{ID}$, creating a tuple $(N_t, N_k)$. 
The logs are then assigned to the group with the same tuple.
This process is based on two assumptions: (1) identifiers from the same template have the same token number; (2) identifiers from the same template have the same key number.
The first assumption is commonly used and has been verified in previous research \cite{iplom, drain}, where the identifiers of event logs are themselves. 

\subsubsection{Fine-grained Grouping}

To further group the logs, \methodname uses a multinomial search tree with a unique search rule for each tuple $(N_t, N_k)$. 
First, the tree forks based on the keys in the $N_k$ layer, and then forks based on non-key prefix tokens in the next $N_p$ ($N_p = N_t - N_k$) layer. 
The tree is initialized with only one root node. 
\methodname creates a new node each time when a new identifier $S_{ID}$ comes and no corresponding node exists at the current layer.
This forking strategy is based on the assumption that template tokens are often at the beginning and variable tokens are often at the end of logs.
To further improve the accuracy of parsing, \methodname allows setting limits on tree depth and identifier length.
This helps prevent too long or too short logs from affecting the parsing accuracy by creating incorrect parsing rules.
Identifiers not meeting the length criteria will be assigned to a designated group for similarity matching, rather than going through the tree.

\subsection{\modify{Online Parsing}}\label{sec:updating}

\modify{Upon a $S_{ID}$ reaching a leaf node, \methodname performs \textit{online parsing}, which involves obtaining current log's template and updating the outdated template in each log group.
It assigns the current log to a specific log group, obtains the group template $S_T$ (\ie the common sequence of logs' $S_{ID}$ in the group) as the current log's template, and dynamically updates $S_T$ with new incoming $S_{ID}$. 
Previous studies~\cite{drain, lenma, iplom} commonly employed automatic template updates and maintenance, which may be ineffective for parsing logs with ambiguous tokens, as discussed in Sec.~\ref{sec:ambiguous}.
Therefore, \methodname introduces a \textit{guided updating} mechanism to enhance parsing performance for log files with ambiguous logs, while retaining the classic \textit{auto updating} mechanism for regular log files. 
The following sections provide a detailed explanation of the two updating methods.
}


\subsubsection{Auto Updating}

\modify{\methodname calculates the similarity (Eq.~\ref{eq:1}) between $S_{ID}$ and all groups' templates within current node.
If a template $S_T$ has the highest similarity score that exceeds the merging similarity threshold $\varepsilon_m$ (\ie a hyperparameter determined before parsing), \methodname will update the template to be the common sequence of $S_{ID}$ and the original template $S_T$ (\ie $S_T \leftarrow S_{ID} \cap S_T$). 
If no such templates exist, \methodname creates a new template and initializes it as $S_{ID}$.
By continuously updating and outputting log templates based on the newly added $S_{ID}$, \methodname achieves log parsing in an online manner.
}

\subsubsection{Guided Updating}\label{sec:guidedupdating}


\modify{
\methodname still calculates the similarity (Eq.~\ref{eq:1}) between $S_{ID}$ and all templates in node. 
However, instead of solely relying on $\varepsilon_m$ to perform updating, \methodname introduces a \textit{"merge-reject"} strategy that allows users to reject a template merge on a potentially ambiguous token in candidate templates.
Specifically, \methodname first conducts a check: \textit{whether tokens requiring updating in $S_T$ contain non-key tokens?}
If yes, it triggers a merge query, \ie output current $S_T$ and $S_{ID}$, and prompts the user to decide whether to reject the automatic merging and updating process. 
If rejected, \methodname creates a new group under the same leaf node, initializing it with $S_{ID}$ as the template. 
Otherwise, the merging process continues, and $S_T$ is updated with the common sequence of tokens between $S_T$ and $S_{ID}$.
The two cases (\ie trigger a merge query or not) are illustrated with examples in Fig.~\ref{fig:feedbackdetails}.}

The idea of developing a human feedback mechanism was inspired by a recent parser, SPINE~\cite{spine}. 
However, \methodname provides a different and more efficient mechanism.
\modify{Specifically, \methodname largely reduces the frequency of triggered queries by skipping the updates of keys.
This is because the keys are casted parameter tokens in the preprocessing and the ambiguous tokens are non-key tokens.
In addition, \methodname does not involve numerical computation to determine whether a merging process should trigger a query in guided mode, making it efficient in development scenarios.}

%% file: Sections/5_Experiments.tex
\section{Experiments}

We conducted extensive experiments on public datasets and industrial datasets to answer the following research questions:
\begin{itemize}
    \item \textbf{RQ1:} How accurate and efficient is \methodname on hybrid logs?
    \item \textbf{RQ2:} How accurate and efficient is \methodname on single-line logs?
    \item \textbf{RQ3:} How effective is \methodname's human feedback integration mechanism?
\end{itemize}
\subsection{Experiment Setup}

\subsubsection{\modify{Implementation}}
All of the experiments were performed on a virtual machine with 128 Intel(R) Xeon(R) Platinum 8375C CPU @ 2.90GHz processors and 94GB RAM on Ubuntu 20.04.5 LTS. 
We implemented \methodname in Python 3.9.12 with the same key casting table shown in Fig.~\ref{fig:flowexample}.
\modify{
Additionally, we implemented specific hyperparameter configuration files to manage all \methodname's hyperparameters for each log source. 
Specifically, the hyperparameters include (1) the maximum tree depth, (2) the maximum and minimum identifiers length, (3) the aggregating similarity threshold $\varepsilon_a$, and (4) the merging similarity threshold $\varepsilon_m$.
Similar to previous works~\cite{drain,spell,spine}, hyperparameters are only related to the log file source and remain unchanged during the parsing process after being determined by grid search, a common hyperparameter tuning method, on small-scale homogeneous log data (\ie 100 random samples for each dataset).}

\modify{
Notably, in experiments, we observed that setting the $\varepsilon_m$ under the guided updating mode (denoted as $\varepsilon_{mg}$) to be approximately $0.2$ lower than $\varepsilon_m$ under the auto updating mode (denoted as $\varepsilon_{ma}$) will result in better parsing performance for most log sources. 
To avoid frequent editing of configuration files when switching updating modes, we introduced an elastic variable $\varepsilon_e = \varepsilon_{ma} - \varepsilon_{mg}$ to optimize the performance in both modes. 
Specifically, $\varepsilon_e$ is typically set to $0.2$ by default. 
Only in a small amount of log sources, $\varepsilon_e$ need to be changed to further enhance parsing performance.
}
\subsubsection{Datasets} 
Our experiments are conducted on 19 log datasets, including 16 single-line log datasets from LogPAI \cite{benchmark} and 3 hybrid log datasets collected by us. 
The LogPAI datasets cover a variety of log types, including OS logs, app logs, and microserver logs, with 2k messages each and 5 to 200 templates. 
The hybrid datasets consist of runtime logs generated by a cloud system benchmark (HiBench), multi-source logs collected by a cloud testing system (CTS), and multi-terminal logs collected in PaaS product in Huawei Cloud (Paas). 
We manually labeled all hybrid log datasets and released the first two datasets for replication. 
Table.~\ref{table:dataset} presents the statistics of these datasets. 
All of these datasets include ground truth labels.

\begin{table}[htbp]\label{table:dataset}
\caption{Dataset statistics of hybrid log datasets.}
\centering
\scalebox{0.9}{
\begin{tabular}{cccc}
    \toprule
     Num. (Event/Table/Text) & HiBench & CTS & PaaS  \\
     \midrule
     Message Num. & 1879/2057/64 &  260/17/101 & 386/36/255  \\
     Template Num. & 92/7/18& 93/7/43&  129/6/93   \\
     \bottomrule
\end{tabular}
}
\label{table:dataset}
\end{table}

\subsubsection{Metrics}

Our evaluation uses two commonly used metrics: Grouping Accuracy \cite{benchmark, guideline, logram} and Template F1-score \cite{guideline}. 
The former measures the proportion of correctly parsed log messages, while the latter measures the F1 score of the generated templates from the parser. 
We use these metrics to conduct fair comparisons at both the message and template levels. 
Additionally, we compare the execution time of different parsers to evaluate their efficiency.

\subsubsection{Baselines}\label{sec:baselines}

We selected AEL \cite{ael}, LenMa \cite{lenma}, Spell \cite{spell}, IPLoM \cite{iplom}, Drain \cite{drain}, and SPINE \cite{spine}, the top-performing unsupervised log parsers, as our baselines for comparison.
\modify{Since \methodname is the first attempt for parsing hybrid logs, there is no other baseline for comparison on parsing multi-line logs.
To demonstrate \methodname's effectiveness on hybrid logs, we collect and flatten multiple lines into individual log entries with line delimiters \texttt{"/n"}, which is a typical solution for managing multi-line logs in real development scenarios~\cite{elastic,flatten1,flatten2,fluentd} (including in Huawei Cloud) before inputting hybrid logs into existing parsers.}
Note that since SPINE is not open-source, we cannot reproduce their results. To improve the scientific rigor of our research, we compare \methodname with SPINE in part of our experiments by using results from its original paper under the same experimental settings, \ie grouping accuracy on single-line logs and feedback results on the Linux dataset.
All baselines are implementations provided by LogPAI or their original repositories with all parameters set to their optimal configuration.

\subsection{RQ1: How Accurate and Efficient is \methodname on Hybrid Logs?}\label{sec:rq1}

For RQ1, we use grouping accuracy, template F1-score, and execution time as evaluation metrics. 
We conducted experiments on 3 hybrid log datasets.
Note that the comparison is not \textit{apple to apple} because \modify{as to our knowledge, Hue is the first parser attempting to parse hybrid logs (including multi-line logs). 
Therefore, we cannot find other more suitable baselines for comparison. Our purpose is not to claim this is a weakness of existing single-line parsers because they were not designed for hybrid log parsing, but rather to highlight the importance of designing a new parser for parsing hybrid logs.
In addition, the experiments intended to show that it is non-trivial to adapt single-line parsers to hybrid logs.}
To be fair, we use the same set of regexes for all parsers including \methodname.

\subsubsection{Accuracy}

\begin{table}[tbp]
\caption{Grouping accuracy on hybrid log datasets.}
\centering
\scalebox{0.95}{
\begin{tabular}{ccccccc}
    \toprule
     Metrics & AEL & LenMa & Spell & IPLoM & Drain & \methodname     \\
     \midrule
     HiBench & 0.308 & 0.238 & 0.360 & 0.300 & 0.442 & \textbf{0.932*}\\
     CTS & 0.620 & 0.432 & 0.571 & 0.513 & 0.746 & 0.848* \\
     PaaS &  0.178 & 0.300 & 0.130 & 0.345 & 0.502 & 0.754* \\
     Average & 0.369 & 0.323 & 0.354 & 0.386 & 0.563 & 0.845* \\
     \bottomrule
\end{tabular}
}
\label{table:hybridga}
\end{table}

\begin{table}[tbp]
\caption{Template F1-score on hybrid log datasets.}
\centering
\scalebox{0.95}{
\begin{tabular}{ccccccc}
    \toprule
     Metrics & AEL & LenMa & Spell & IPLoM & Drain & \methodname     \\
     \midrule
     HiBench  & 0.083 & 0.590 & 0.641 & 0.687 & 0.753 & \textbf{0.836*}\\
     CTS & 0.662 & 0.569 & 0.618 & 0.645 & 0.784 & \textbf{0.839*} \\
     PaaS  &  0.326 & 0.388 & 0.292 & 0.450 & 0.540 & \textbf{ 0.801*} \\
     Average & 0.357 & 0.516 & 0.517 & 0.594 & 0.692 & \textbf{0.825*} \\
     \bottomrule
\end{tabular}
}
\label{table:hybridf1}
\end{table}

The results are shown in Table.~\ref{table:hybridga} and Table.~\ref{table:hybridf1}. 
The best results in each dataset are marked with a star symbol, and the results with grouping accuracy greater than 0.9 and F1-score greater than 0.8 are emphasized in bold font.
\methodname demonstrates state-of-the-art accuracy and F1-score on all three hybrid log datasets, outperforming the best traditional parser, Drain. 
In comparison, Drain achieves an average grouping accuracy of 0.563 and an average F1-score of 0.692, while \methodname achieves the highest average grouping accuracy of 0.845 and the highest average F1-score of 0.825.
This superior performance can be attributed to the design of \methodname, which is specifically built to handle the unique challenges posed by hybrid log parsing. 
Unlike traditional parsers, \methodname is able to effectively parse logs that span multiple lines and retain the structure between lines, whereas traditional parsers fall short because they were not designed to natively support multi-line log parsing.
As mentioned in Sec.~\ref{sec:neglected}, the structure between lines is lost after flattening logs across lines. 
For example, in table logs, different instances of the same template may have different line numbers, which can result in significant differences in the message lengths after flattening. 
This, in turn, results in far more templates than ground truth, making it difficult or even impossible for downstream tasks to use the logs for vectorized representation.

\begin{figure}[htpb]
\centering
\includegraphics[scale=0.185]{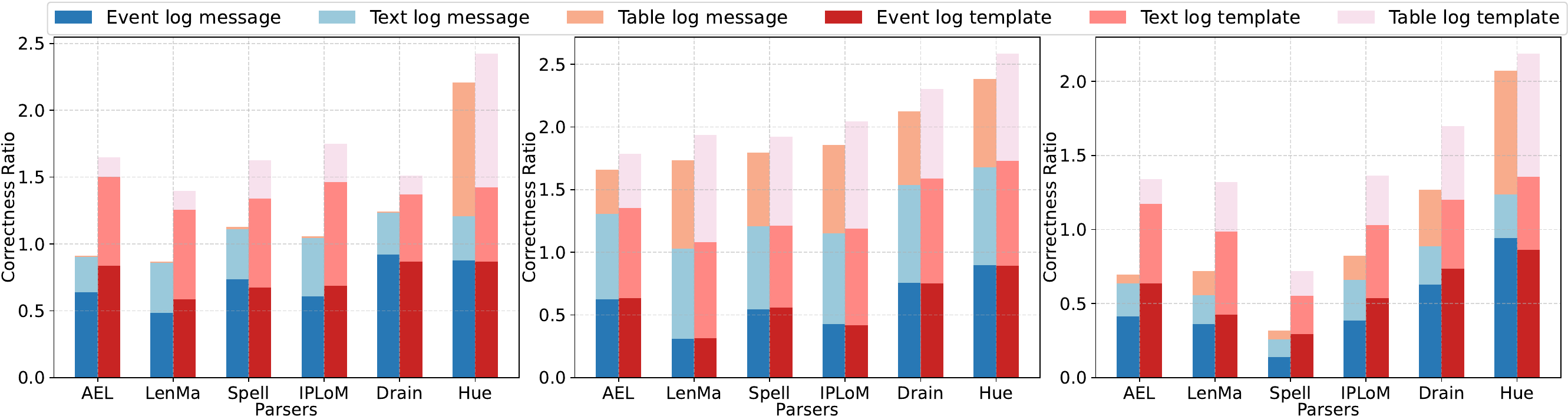}   
\caption{
The ratio of correctly parsed message/template numbers in each type.
The upper bound of each stacked bar is 3.0.}
\label{fig:composition}
\end{figure}

\modify{In particular, we find that \methodname has a significant advantage in parsing event logs and table logs.
Fig.~\ref{fig:composition} illustrates the comparative advantages, \ie calculating the ratio of correctly parsed message number or template number dividing the total message number or template number on each log type when they are parameterized to maximize template F1-score.
For example, on HiBench, \methodname demonstrates promising performance in parsing table logs, while other log parsers struggle to parse table logs. 
In addition, on CTS, \methodname exhibits significant advantages in parsing event logs.
These results demonstrate the necessity of designing a parser specifically for hybrid logs, and migrating existing parsers for hybrid log parsing is not trivial.}



\begin{figure}[htbp]
\centering
\includegraphics[scale=0.185]{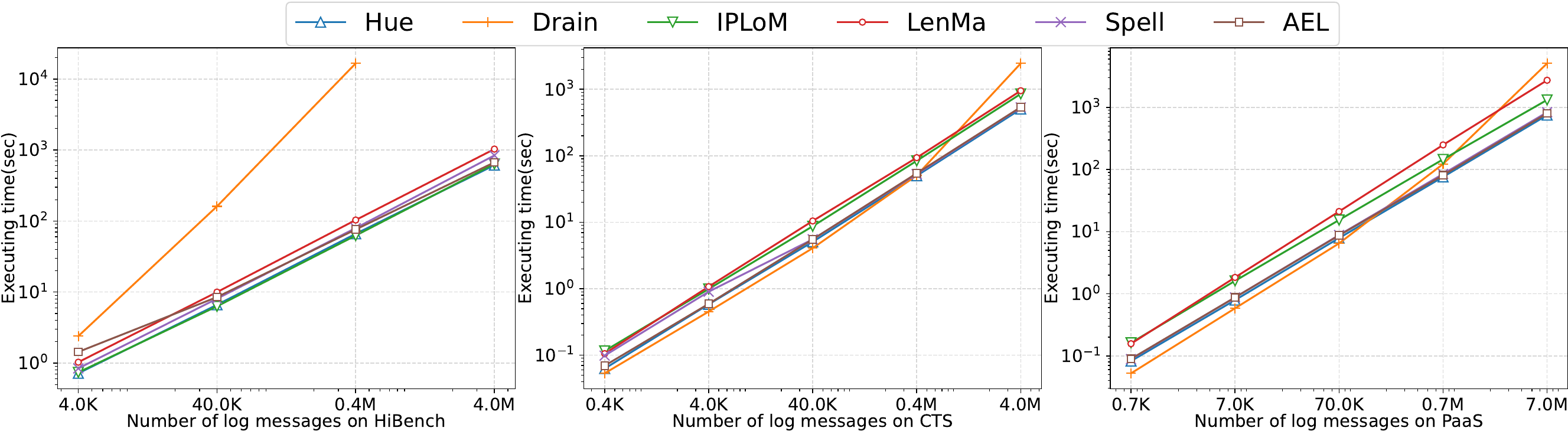}   
\caption{Execution time on hybrid logs.}
\label{fig:exectimehybrid}
\end{figure}

\subsubsection{Efficiency}

\modify{We specially collected a large quantity of log data on these three hybrid logs for efficiency evaluation.
Specifically, we compare average time consumption after running five times in a scaling-up scenario.
}
The results of the experiments are displayed in Fig.~\ref{fig:exectimehybrid}.
We observe that \methodname obtains SOTA results and its execution time increases linearly with the number of logs.
Some multi-line logs, after being flattened, can be very long and cause sequence clustering-based parsers to fail.
\modify{Although Drain performs the best effectiveness among all baselines, it still lacks of efficiency on hybrid logs.
Specifically, Drain times out when the number of messages reached 4.0 million.}
The reason might be that Drain relies on direct comparison to group extra-long logs, which is highly inefficient on flattened multi-line logs.

\subsection{RQ2: How Accurate and Efficient is \methodname on Single-line Logs?}\label{sec:rq2}

For RQ2, we use grouping accuracy, template F1-score, and execution time for evaluation. 
We compared \methodname (auto-updating) with 6 SOTA log parsers on 16 open-source single-line log datasets.

\subsubsection{Accuracy}

\begin{table}[tbp]
\caption{Grouping accuracy on LogPAI's loghub datasets.}
\centering
\scalebox{0.85}{
\begin{tabular}{cccccccc}
    \toprule
     Dataset & AEL & LenMa & Spell & IPLoM & Drain & SPINE & \methodname     \\
     \midrule
     HDFS & \textbf{0.998} & \textbf{0.998} & \textbf{1*} & \textbf{1*} & \textbf{0.998} & \textbf{0.998}  & \textbf{0.998}   \\ 
     Spark & \textbf{0.905} & 0.884 & \textbf{0.905} & \textbf{0.920} & \textbf{0.920} & \textbf{0.925} & \textbf{0.942*}\\
     BGL & 0.758 & 0.690 & 0.787 & \textbf{0.939} & \textbf{0.963*} & \textbf{0.948} & 0.849\\
     Windows & 0.690 & 0.566 & \textbf{0.989} & 0.567 & \textbf{0.997*} & \textbf{0.990} & \textbf{0.990}\\
     Linux & 0.690 & 0.701 & 0.605 & 0.672 & 0.690 & 0.676 & 0.749*\\
     Android & 0.682 & 0.880 & \textbf{0.919} & 0.712 & \textbf{0.911} & \textbf{0.932*} & 0.826\\
     Mac & 0.764 & 0.698 & 0.757 & 0.673 & 0.787 & 0.789 & \textbf{0.901*}\\
     Hadoop & 0.538 & 0.885 & 0.778 & \textbf{0.954} & \textbf{0.948} & \textbf{0.946} & \textbf{0.966*}\\
     HealthApp & 0.568 & 0.174 & 0.639 & 0.822 & 0.780 & \textbf{0.988*} & \textbf{0.903} \\
     OpenSSH & 0.538 & \textbf{0.925*} & 0.554 & 0.802 & 0.788 & 0.681 & 0.804\\
     Thunderb. &  \textbf{0.941} & \textbf{0.943} & 0.844 & 0.663 & \textbf{0.955} & \textbf{0.964*} & \textbf{0.962}\\
     Proxifier & 0.518 & 0.508 & 0.527 & 0.515 & 0.527 & \textbf{0.967} & \textbf{1*}\\
     Apache & \textbf{1*} & \textbf{1*} & \textbf{1*} & \textbf{1*} & \textbf{1*} & \textbf{1*} & \textbf{1*}\\
     HPC &  \textbf{0.903} & 0.830 & 0.654 & 0.824 & 0.887 & 0.871 & \textbf{0.951*} \\
     Zookeeper & \textbf{0.921} & 0.841 & \textbf{0.964} & \textbf{0.962} & \textbf{0.967} & \textbf{0.989*} & \textbf{0.987} \\
     OpenStack & 0.758 & 0.743 & 0.764 & 0.871 & 0.733 & 0.757 & \textbf{0.993*} \\
     Average &0.754 &0.721 &0.751 &0.777 &0.865 & \textbf{0.901} & \textbf{0.927*}\\
     \bottomrule
\end{tabular}
}
\label{table:singlega}
\end{table}
\begin{table}[tbp]
\caption{Template F1-score on LogPAI's loghub datasets.}
\centering
\scalebox{0.95}{
\begin{tabular}{cccccccc}
    \toprule
     Dataset & AEL & LenMa & Spell & IPLoM & Drain & \methodname     \\
     \midrule
     HDFS  & \textbf{0.998} & \textbf{0.867} & \textbf{1*} & \textbf{1*} & \textbf{0.867} & \textbf{0.867}   \\
     Spark  & 0.511 & 0.373 & 0.510 & 0.769 & 0.780 & \textbf{0.870*}\\
     BGL & \textbf{0.819} & 0.412 & 0.487 & \textbf{0.832*} & \textbf{0.830} & 0.700\\
     Windows  & \textbf{0.833} & 0.752 & 0.777 & 0.774 & \textbf{0.932*} & \textbf{0.811}\\
     Linux  & \textbf{0.818} & \textbf{0.944*} & \textbf{0.802} & 0.796 & \textbf{0.930} & \textbf{0.926}\\
     Android  & 0.726 & \textbf{0.838} & \textbf{0.867*} & 0.703 & \textbf{0.825} & \textbf{0.834}\\
     Mac  & 0.759 & 0.655 & 0.634 & 0.743 & 0.797 & \textbf{0.885*}\\
     Hadoop  & 0.702 & 0.672 & 0.604 & \textbf{0.863} & \textbf{0.820} & \textbf{0.880*}\\
     HealthApp  & 0.306 & 0.114 & 0.367 & 0.463 & 0.351 & \textbf{0.960*}\\
     OpenSSH & 0.571 & \textbf{0.824*} & 0.630 & 0.667 & \textbf{0.824*} & 0.767\\
     Thunderb.  & 0.695 & 0.748 & 0.638 & 0.784* & 0.773 & 0.779\\
     Proxifier & 0.400 & 0.154 & 0.134 & 0.600 & 0.534 & \textbf{1*}\\
     Apache  & \textbf{1*} & \textbf{1*} & \textbf{1*} & \textbf{1*} & \textbf{1*} & \textbf{1*}\\
     HPC  & 0.750 & 0.297 & 0.739 & 0.711 & 0.753 & \textbf{0.872*} \\
     Zookeeper  & 0.718 & 0.755 & 0.729 & 0.744 & \textbf{0.854*} & \textbf{0.854*} \\
     OpenStack  & 0.657 & 0.244 & 0.089 & 0.693 & 0.117 & \textbf{0.978*}\\
     Average & 0.696& 0.603 & 0.629 & 0.759 & 0.749 & \textbf{0.870*}\\
     \bottomrule
\end{tabular}
}
\label{table:singlef1}
\end{table}

The results are presented in Table~\ref{table:singlega} and Table~\ref{table:singlef1}.
\methodname achieves the SOTA results with the highest average values in both grouping accuracy and template F1-score. 
SPINE \cite{spine} is the best among existing parsers on single-line logs.
Compared with SPINE, \methodname performs better than at the message level (Grouping accuracy) on 8 datasets and is comparable on 4 datasets. 
\methodname also achieves SOTA results at the template level for 9 datasets.
To sum up, \methodname achieves the highest average values of 0.927 in grouping accuracy and 0.870 in template F1-score, surpassing the best traditional parser results of 0.901 and 0.749, respectively.

We also observe that \methodname's accuracy and F1-score are lower on certain datasets (\eg Linux). 
We think this is caused by some ambiguous tokens in the logs.
For instance, \texttt{"usbcore: registered new driver} \textit{usbfs}\texttt{"} and \texttt{"usbcore: registered new driver} \textit{hub}\texttt{"} result in lower accuracy on this dataset. 
To address this issue, \methodname provides the human feedback integration mechanism (guided updating), which will be discussed in the following section.

\subsubsection{Efficiency}

\begin{figure}[tpb]
\centering
\includegraphics[scale=0.185]{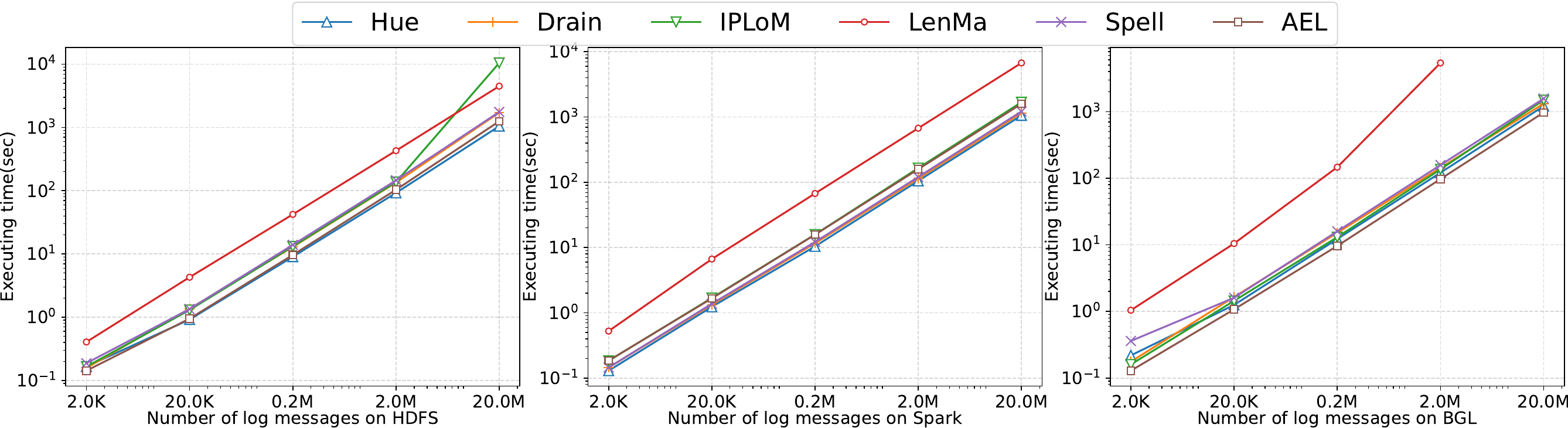}   
\caption{Execution time on single-line logs.}
\label{fig:exectimesingle}
\end{figure}

\modify{We assessed the efficiency the parsers on three major datasets, \ie using HDFS, Spark, and BGL for executing time comparisons~\cite{spine}. 
We also compare average time consumption after running five times in a scaling-up scenario.
}
The results are shown in Fig.~\ref{fig:exectimesingle}.
The results indicate that \methodname is comparable with the best baseline in terms of processing speed for single-line log data.
However, \methodname is slightly slower than the fastst baseline in some cases, which we argue is acceptable.



\subsection{RQ3: How Effective is \methodname's Human Feedback Integration Mechanism?}\label{sec:rq3}

\modify{For RQ3, we conducted experiments on all three hybrid logs as well as the three single-line logs that achieve relatively unsatisfactory grouping accuracy (below 0.9) and consist of a large number of log templates (over 100 templates) in Sec.~\ref{sec:rq2} (\ie Linux, Android, and BGL) to highlight the improvement brought by \methodname's feedback mechanism.}
\modify{Specifically, we set \methodname to \textit{guided updating} mode and evaluate the improvement on various datasets by changing the maximum limit of triggered merge queries. 
Furthermore, in the experiments, we keep all feedback provided by the user correct to eliminate any potential perturbation caused by human factors.}

\begin{figure*}[tpb]
\centering
\begin{subfigure}[t]{0.33\linewidth}
\includegraphics[scale=0.35]{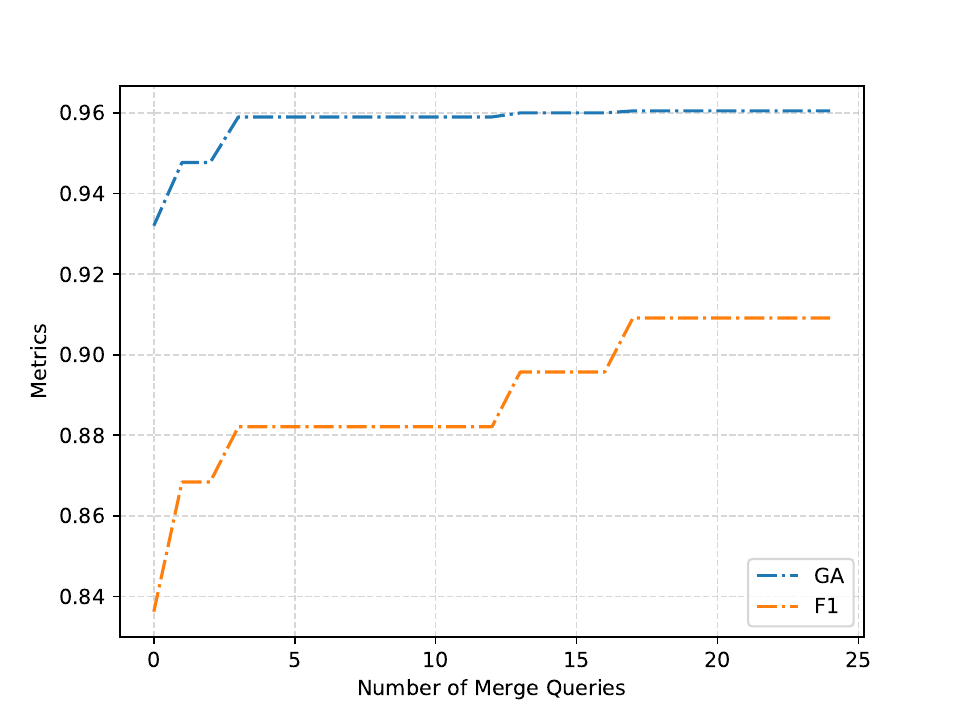}   
\caption{HiBench}
\end{subfigure}
\begin{subfigure}[t]{0.33\linewidth}
\includegraphics[scale=0.35]{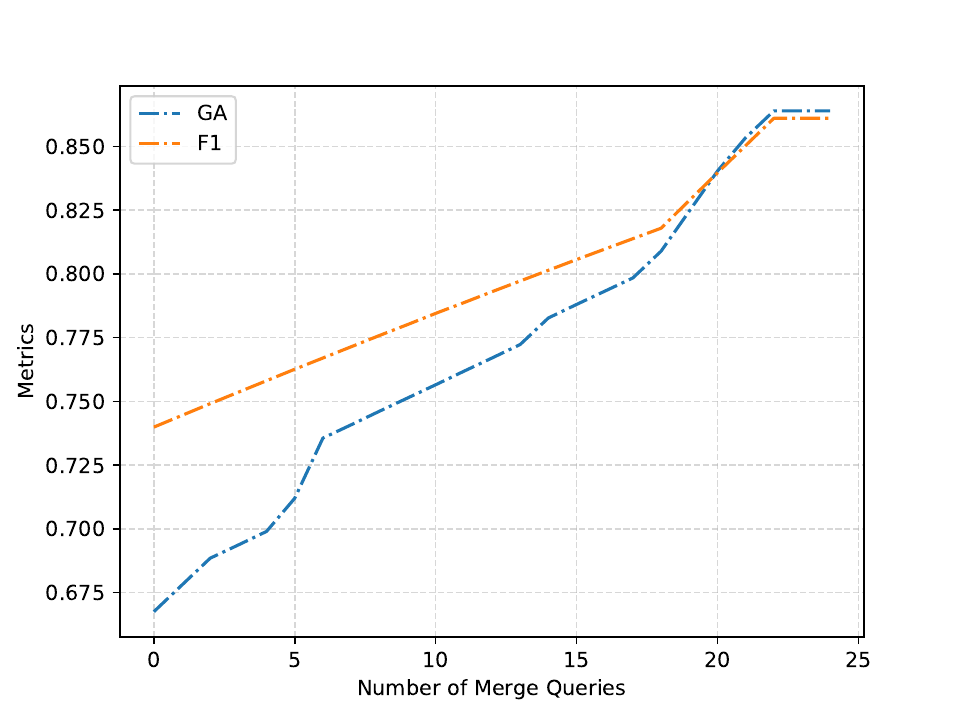}   
\caption{CTS}
\end{subfigure}
\begin{subfigure}[t]{0.33\linewidth}
\includegraphics[scale=0.35]{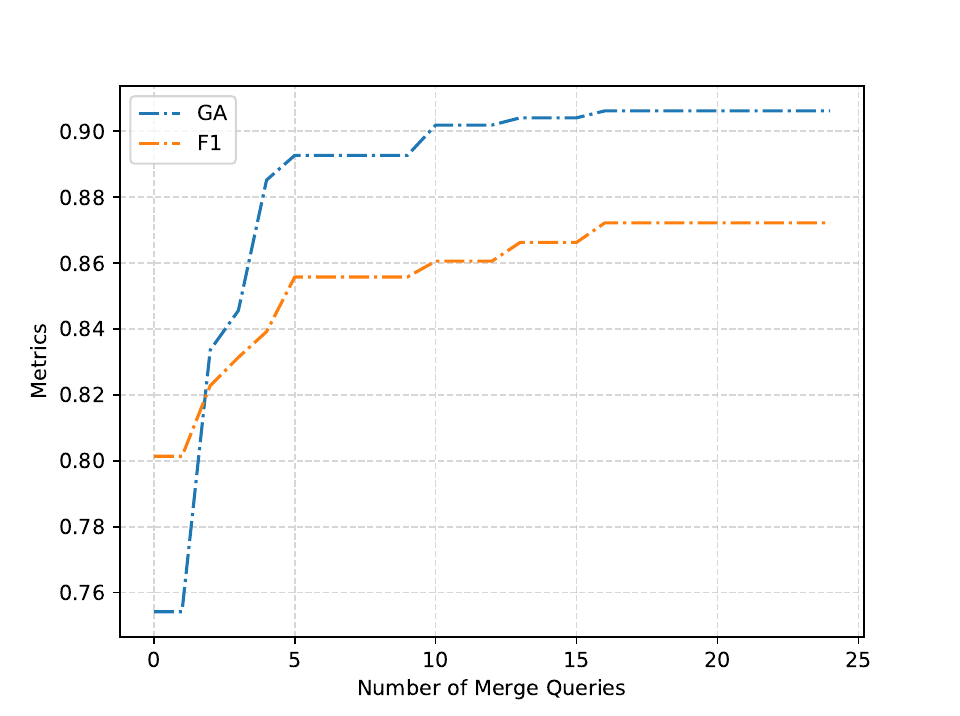}   
\caption{PaaS}
\end{subfigure}
\caption{The increase of grouping accuracy and template F1-score on hybrid log datasets with feedback.}
\label{fig:feedbackhybrid}
\end{figure*}

\begin{figure*}[tpb]
\centering
\begin{subfigure}[t]{0.33\linewidth}
\includegraphics[scale=0.35]{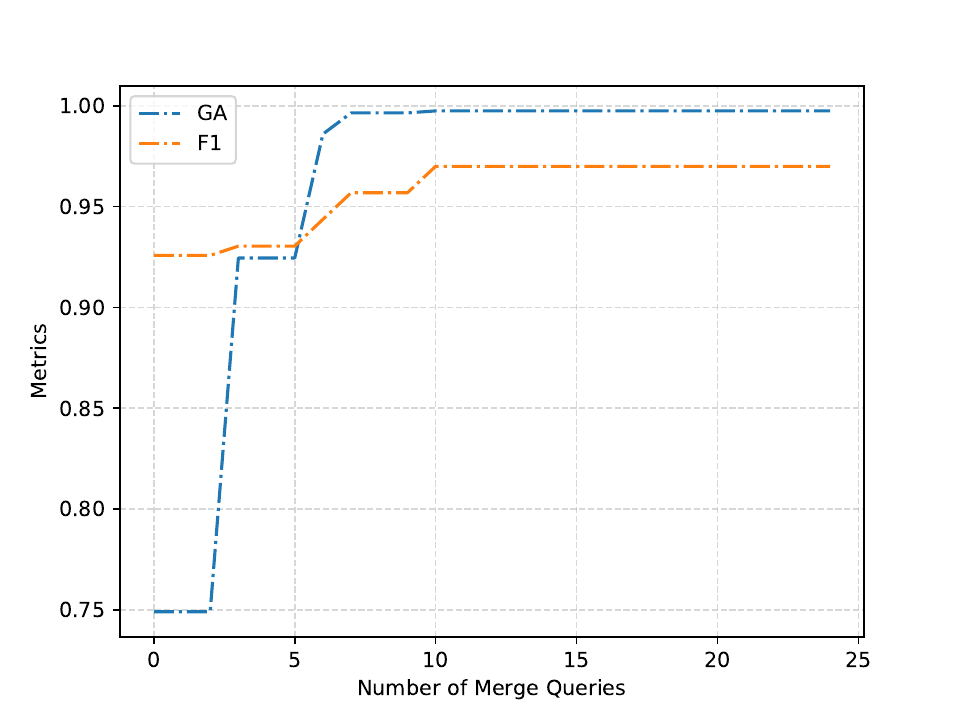}   
\caption{Linux}
\end{subfigure}
\begin{subfigure}[t]{0.33\linewidth}
\includegraphics[scale=0.35]{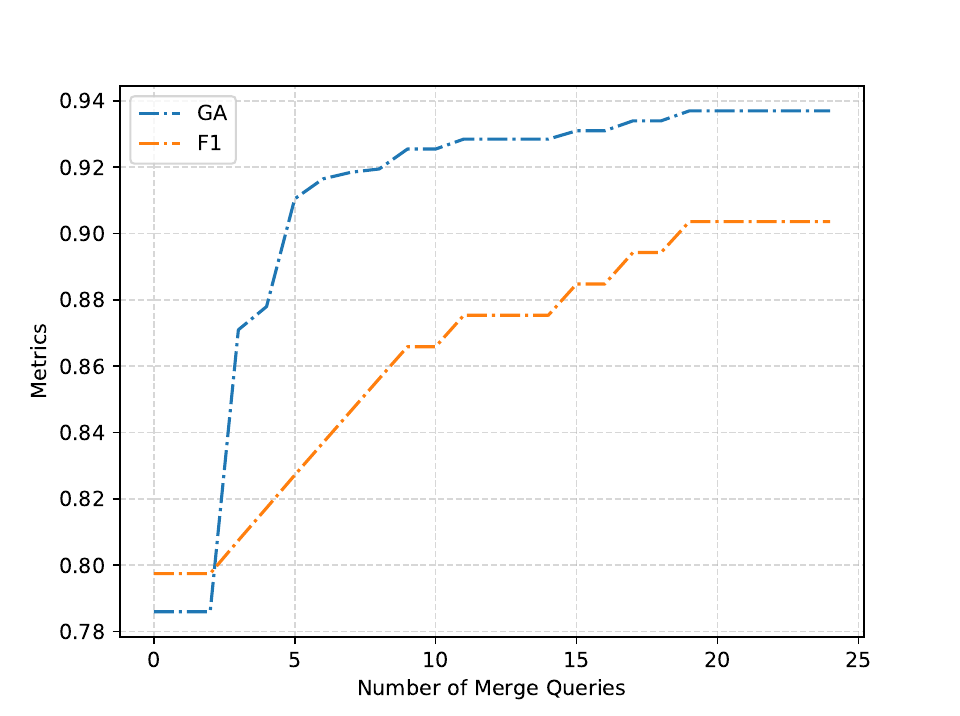}   
\caption{Android}
\end{subfigure}
\begin{subfigure}[t]{0.33\linewidth}
\includegraphics[scale=0.35]{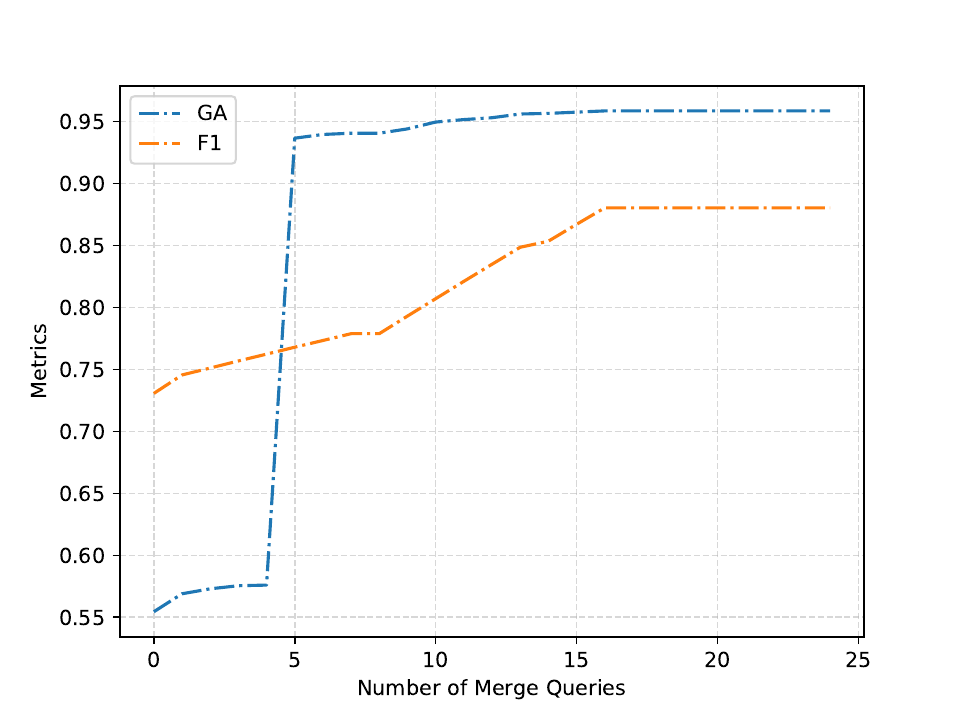}   
\caption{BGL}
\end{subfigure}
\caption{The increase of grouping accuracy and template F1-score on three LogPAI datasets with feedback.}
\label{fig:feedbacksingle}
\end{figure*}

The results are shown in Fig.~\ref{fig:feedbackhybrid} and Fig.~\ref{fig:feedbacksingle}.
Aside from CTS, \methodname's grouping accuracy was able to improve to over 0.92 and template F1-score to over 0.87 for all datasets with less than 20 feedback queries. 
The performance on CTS might not be optimal, which could be due to the relatively high number of templates and low amount of corresponding logs in the dataset.

\begin{figure}[thpb]
\centering
\includegraphics[scale=0.35]{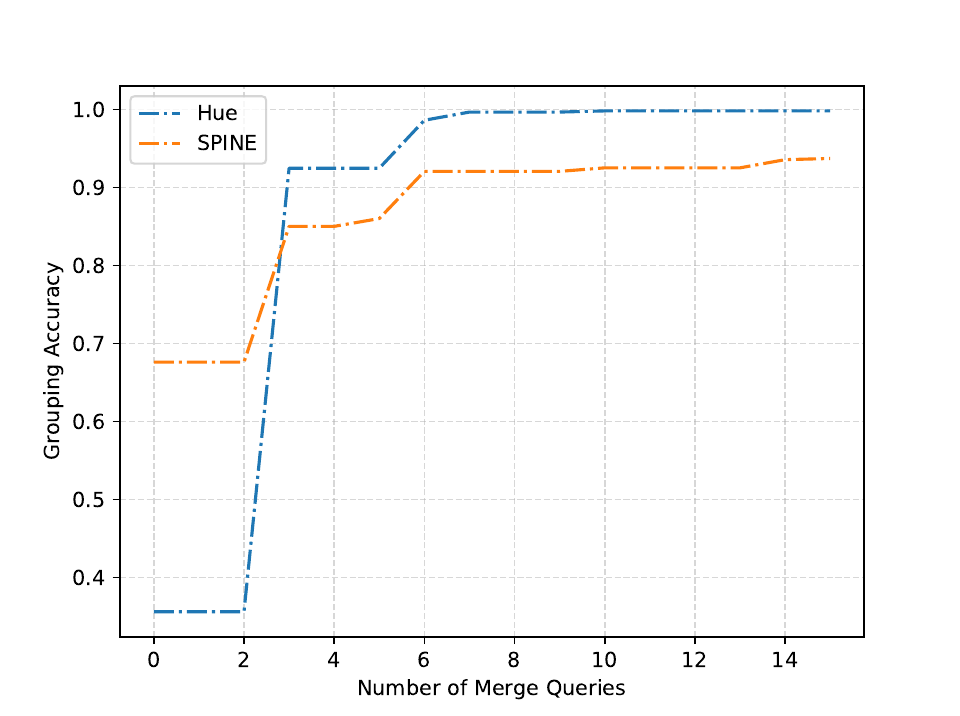}   
\caption{Feedback mechanism comparison on Linux.}
\label{fig:feedbackcompare}
\end{figure}

For effectiveness comparison, we selected SPINE-Feedback~\cite{spine} to demonstrate the superiority of \methodname's feedback integration mechanism.
\modify{We carried out experiments on Linux where both \methodname and SPINE did not perform satisfactorily, \ie with the lowest grouping accuracy of 0.749 and 0.676 among all 16 single-line datasets, respectively}.
To highlight the strength of \methodname's feedback mechanism, we intentionally reduced \methodname's base grouping accuracy by tuning its parameters to be lower than SPINE's to highlight its feedback mechanism strength, despite its initially higher accuracy than SPINE's in auto-updating mode.

\modify{The results are shown in Fig.~\ref{fig:feedbackcompare}.}
\modify{Compared with SPINE,  \methodname is able to improve accuracy efficiently with fewer feedback queries significantly.
Moreover, it achieves a higher final grouping accuracy after tens of queries. 
The efficient and effective feedback performance of \methodname may be attributed to two aspects: 
(1) \methodname pre-filters a large number of unnecessary feedback queries by checking whether all tokens to be updated are keys. 
As discussed in Sec.~\ref{sec:guidedupdating}, \methodname assumes that keys are potential parameter tokens or non-ambiguous tokens. 
Therefore, the template updates that are solely targeting keys are considered as correct updates and will not trigger feedback queries.
(2) The feedback mechanism of \methodname resembles \textit{"preventing erroneous template updates,"} whereas the feedback mechanism of SPINE resembles \textit{"correcting errors after they occur."} 
Specifically, SPINE iteratively partitions the already merged groups into new groups and triggers feedback queries each time. 
However, this may not be able to resolve the cases where logs belonging to the same template have already been clustered in the wrong groups, where the partition feedback queries cannot help them correctly clustered together again. 
In contrast, \methodname raises feedback queries before merging logs into wrong groups, which can effectively prevent incorrect grouping cases at the very beginning.
Thus, \methodname can effectively achieve higher final accuracy.
}

Additionally, we observe that \methodname's accuracy finally saturates as the number of merge queries increases in some experiments.
This is because the parsing errors could also come from other components of \methodname (\eg from the heuristic parse tree in fine-grouping).

%% file: Sections/6_Industrial.tex
\section{Industrial Deployment}\label{sec:industry}

We deployed \methodname in Huawei Cloud and it ran smoothly for two months, from October 2022 to December 2022. 
During this period, we annotated 1,000 logs 
for each downstream task every week. 
Throughout the evaluation, IT operators did not have access to the source code of the log statements.

\begin{figure}[htpb]
\centering
\begin{subfigure}[t]{0.47\linewidth}
\includegraphics[scale=0.35]{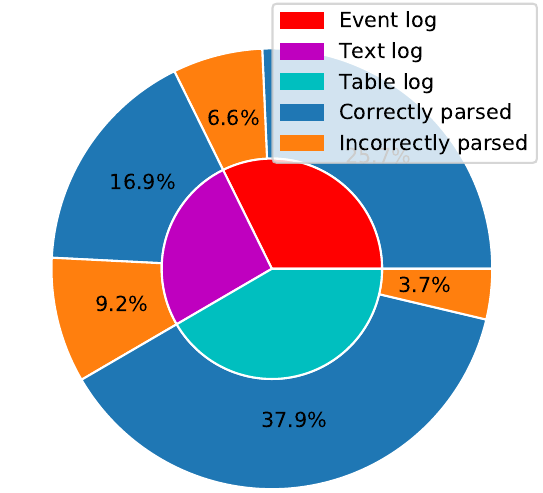}   
\caption{Composition.}
\end{subfigure}
\begin{subfigure}[t]{0.47\linewidth}
\includegraphics[scale=0.25]{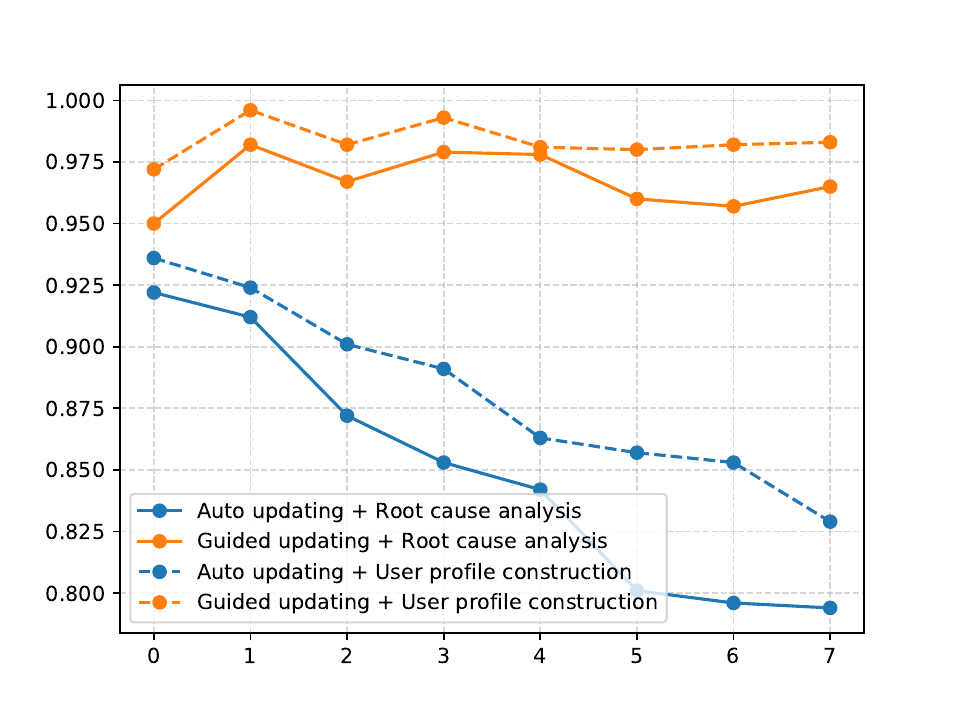}   
\caption{Grouping accuracy.}
\end{subfigure}
\caption{Results in Huawei Cloud.}
\label{fig:industrial}
\end{figure}

To assess the performance of \methodname, we calculated the percentage of each type of log message over the two-month period for root cause analysis tasks.
As shown in Fig.~\ref{fig:industrial}, the blue part represents correctly parsed templates, while the orange part represents incorrectly parsed templates.
Out of all parsed log templates, 32.3\% were event logs, 26.1\% were text logs, and 41.6\% were table logs. Nearly 70\% of them were multi-line, making it challenging for traditional log parsers to provide accurate log templates for downstream tasks.

\modify{
In our department in Huawei Cloud, the logs are used for two downstream tasks: root cause analysis and user profile construction. 
Since IT operators cannot directly access the source code and extract templates from logging statements, they are limited to manually defining log templates according to their requirements.
However, as discussed in Sec.~\ref{sec:ambiguous}, distinct downstream tasks may possess varying interpretations of log templates.
For example, event logs typically exhibit a "verb-object" structure, manifesting as "performed an action on a specific object" (Fig.~\ref{fig:ambiguousexample}).
In the root cause analysis task, engineers focus on the objects being operated, so they partition templates based on objects.
In the user profile construction task, engineers focus on the recorded action, so they partition templates based on actions. 
To achieve more accurate log parsing aligned with their downstream task requirements, they utilized \methodname's guided updating mode to parse the collected log samples with different human feedback.
}

Fig.~\ref{fig:industrial} shows the results, where the solid and dashed lines are for root cause analysis and user profile construction, and blue and orange lines are for auto-updating and guided-updating modes, respectively. 
\modify{This performance has been deemed acceptable by the engineers responsible for their respective downstream tasks. 
They have successfully integrated the parsed hybrid log data into their experimental models.}
\modify{Furthermore, in our discussion with engineers, we learned that \methodname did not trigger an excessive number of feedback queries during the actual parsing process, thus avoiding user fatigue in providing feedback. 
In fact, the average number of feedback queries triggered in every 1,000 logs was less than 30, while the grouping accuracy remained acceptable (\ie over 0.9). 
This enabled them to try to incorporate hybrid logs as available data in their models for the first time. 
We also found that \methodname's space-based tokenization strategy (discussed in Sec.~\ref{sec:keycast}) has minimal impact on the downstream tasks in Huawei Cloud, contrary to our previous belief that it could threaten the practicality of \methodname. 
This is primarily due to the fact that Huawei Cloud's business solely uses vectorized logs for various tasks, where the main requirement is accurate clustering of logs belonging to the same template, rather than ensuring every token in the extracted templates (represented as strings) is correctly parsed.
Therefore, we believe that \methodname also demonstrates a certain level of practicality in industrial scenarios.
}

%% file: Sections/7_Threats.tex
\section{Threats to Validity}
In this work, we identified the following major threats to validity.
\
\begin{itemize}
\item \textbf{Data Quality}: Our evaluation uses multiple log datasets and we found that they contain labeling errors that can negatively impact parser performance. Also, most parsers, including \methodname, assume that logs have a specific header. \methodname segments logs based on headers.
For logs without headers, \methodname's performance would be degraded. 
Thus, we regard parsing hybrid logs without headers as future work.

\item \textbf{Key Adaptability}: In Sec.~\ref{sec:rq2}, we use the most common key setting for all experiments. However, some logs may not fit this setting, which could affect the parsing result. To mitigate this threat, we allow users to customize their own key casting configurations.
    
\item \textbf{Query Correctness}: In Sec.~\ref{sec:rq3}, we assume that human 
 feedback to the query is always correct. However, if the user provided incorrect feedback, it might introduce new parsing errors. 
 To mitigate this impact, we recommend that users discuss the query in detail if they are not sure about the correct template.

\end{itemize}

%% file: Sections/8_Related.tex
\section{Related works}

Log parsing methods can be categorized into unsupervised and supervised methods according to the parsing algorithms. 

\textbf{Unsupervised log parsers} do not require labels on existing log data and thus they have been more widely explored. 
There are three main categories of unsupervised log parsers: frequent pattern mining-based, clustering-based, and heuristic-based methods. 
(1) \textit{Frequent pattern mining-based} methods regard the mined frequent patterns (\eg n-grams) as log templates.
For example, SLCT~\cite{slct}, LFA~\cite{lfa}, LogCluster~\cite{logcluster}, and Logram~\cite{logram} try to use different methods to extract frequent patterns in logs.
(2) \textit{Clustering-based} methods aim to group similar logs first, assuming that logs in the same cluster share the same template, and then extract the common tokens as the template in each cluster.
Some clustering-based methods can perform in an online manner because they adopt an online grouping strategy rather than clustering all the offline logs at once.
Specifically, LKE~\cite{lke}, LogSig~\cite{logsig}, LogMine~\cite{logmine} are offline methods, SHISO~\cite{shiso}, and LenMa~\cite{lenma} are online methods.
(3) \textit{Heuristic-based} methods encode expert domain knowledge into general and effective heuristic rules.
For example, AEL~\cite{ael}, IPLoM~\cite{iplom}, Spell~\cite{spell}, and Drain~\cite{drain} utilize different heuristic rules to extract templates from logs.
In particular, Drain~\cite{drain} achieved SOTA in all open-source traditional parsers with a parse tree structure to perform log parsing in an online manner.
POP~\cite{pop} improves Drain and provides a parallel implementation on Spark for distributed deployment.
SPINE~\cite{spine} improves Drain and proposes a progressive clustering step for human feedback. In particular, SPINE introduced a manual correction paradigm to log parsing.
Although SPINE also considers human feedback, this paper proposed a new human feedback integration mechanism based on the merge-reject strategy, which largely improves efficiency and achieved higher accuracy.

\textbf{Supervised log parsers} are less common because they rely on lots of precise ground truth labels manually constructed by IT operators.
UniParser~\cite{uniparser} utilizes a contrast learning strategy to overcome the pattern difference in heterogeneous log sources and uses a BiLSTM~\cite{bilstm} model for token and context encoding.
SemParser~\cite{semparser} utilizes a special semantic miner for template extracting and uses a BiLSTM~\cite{bilstm} model for encoding.
Different from these supervised approaches, \methodname is an unsupervised log parser that does not rely on label log datasets.

\modify{Different from these existing parsers, \methodname is the first attempt to parse hybrid logs, providing an approach to utilize neglected hybrid logs for various downstream tasks in log analysis.}
\modify{Moreover, most existing parsers can only improve parsing performance through hyperparameter tuning, while \methodname, in addition to hyperparameter tuning, further enhances parsing effectiveness through a feedback mechanism.}


%% file: Sections/9_Conclusion.tex
\section{Conclusion}

This paper introduces a new, general, and practical research problem, hybrid log parsing, which is the superset of the widely-studied single-line log parsing problem.
We also present \textit{\methodname}, the first attempt to tackle this problem. 
\methodname leverages both patterns in the incoming log messages and expert domain knowledge to parse both hybrid and single-line logs in an online manner. 
The use of a key casting table and a merge-reject strategy enables \methodname to effectively utilize user feedback, making it possible to quickly adapt to complex and changing log templates. 
The results of our evaluation on three hybrid log datasets and sixteen widely-used single-line log datasets demonstrate the superiority of \methodname in terms of accuracy and efficiency. 
The successful deployment of \methodname in a real production environment further highlights its practical value.
We hope \methodname and its replication package can serve as the first step in hybrid log parsing and benefit a line of future research in this interesting and practice direction.